\documentclass{Interspeech}

\interspeechcameraready

\usepackage{svg}

\usepackage{amsmath}
\usepackage{soul}
\usepackage[textsize=tiny]{todonotes}

\title{Quantifying and Reducing Speaker Heterogeneity within the Common Voice Corpus for Phonetic Analysis}

\author[affiliation={1}]{Miao}{Zhang}
\author[affiliation={1}]{Aref}{Farhadipour}
\author[affiliation={1}]{Annie}{Baker}
\author[affiliation={1}]{Jiachen}{Ma}
\author[affiliation={1}]{Bogdan}{Pricop}
\author[affiliation={1}]{Eleanor}{Chodroff}

\affiliation{Department of Computational Linguistics}{University of Zurich}{Switzerland}
\email{miao.zhang@uzh.ch, aref.farhadipour@uzh.ch, eleanor.chodroff@uzh.ch}

\usepackage{comment}

\begin{document}

\maketitle

\begin{abstract}
With its crosslinguistic and cross-speaker diversity, the Mozilla Common Voice Corpus (CV) has been a valuable resource for multilingual speech technology and holds tremendous potential for research in crosslinguistic phonetics and speech sciences. Properly accounting for speaker variation is, however, key to the theoretical and statistical bases of speech research. While CV provides a client ID as an approximation to a speaker ID, multiple speakers can contribute under the same ID. This study aims to quantify and reduce heterogeneity in the client ID for a better approximation of a true, though still anonymous speaker ID. Using ResNet-based voice embeddings, we obtained a similarity score among recordings with the same client ID, then implemented a speaker discrimination task to identify an optimal threshold for reducing perceived speaker heterogeneity. These results have major downstream applications for phonetic analysis and the development of speaker-based speech technology.\footnote{\url{https://github.com/pacscilab/CV_clientID_cleaning}}
\end{abstract}

\section{Introduction}
As a massively multilingual spoken corpus, the Mozilla Common Voice Corpus has been pivotal in the development of novel and advanced speech technologies and holds tremendous potential for advancing research in crosslinguistic phonetics and speech sciences \cite{ardila-etal-2020-common, ahn2022voxcommunis}. Properly accounting for speaker variation is, however, key to the theoretical and statistical bases of speech research. While Common Voice provides a client ID as an approximation to a speaker ID, multiple speakers can contribute under the same ID. This study aims to quantify and reduce heterogeneity in the client ID to better approximate a true, though still anonymous, speaker ID. Using a ResNet-293 speaker verification system, we obtained a cosine similarity score among recordings with the same client ID and implemented a speaker discrimination task to identify an optimal threshold for reducing perceived speaker heterogeneity. These results have major downstream applications for phonetic analysis and the development of speaker-based speech technology.

For large-scale crosslinguistic phonetic analysis, Common Voice has several advantages compared to other multilingual speech corpora, notably in its simultaneous breadth and depth. Version 21 of the corpus has over 130 languages, demonstrating substantial crosslinguistic breadth and reasonable diversity (cf. \cite{pratap20_interspeech, wang-etal-2021-voxpopuli}). In Common Voice, each recording is additionally accompanied by a corresponding transcript (cf., \cite{voxlingua2021} and \cite{li2023yodas} with only partially labeled data), dozens to thousands of speakers per language (cf., \cite{ladefoged2009, black2019cmu, conneau2023fleurs} with only one or few speakers per language), and in several cases, a sizable amount of data per speaker (cf., \cite{ladefoged2009, chodroff-etal-2024-phonetic} with only a few minutes of data per speaker). Many of these attributes are also useful for crosslinguistic phonetic analysis: the presence of a transcript is beneficial for the automated corpus phonetics pipeline, in which the audio recordings can be readily converted to a phonetic format for downstream acoustic-phonetic analysis. Speaker diversity can also facilitate better estimates of norms within a language community for improved between-language comparisons.

Common Voice is fully crowdsourced and its ease of accessibility for contributors has likely facilitated its success as a truly massively multilingual speech corpus.
However, due to the crowdsourced data collection procedure, speaker heterogeneity within a corresponding ``client ID'' has been a known issue of Common Voice \cite{hintz2024commonbench}. The ``client ID'' is an identifier released with each recording from Common Voice that ``refers to an anonymized ID for the speaker'' \cite[p. 4219]{ardila-etal-2020-common}. 
Indeed, it is common for Common Voice users to use the client ID as the best approximation of a speaker ID \cite{alam2022bengali, aepli-etal-2022-findings, armentano2024becoming}. 
Recordings linked to a specific client ID may involve several speakers, potentially reducing the speaker count. Conversely, recordings from different devices or sessions without an account might each get a unique client ID, possibly inflating the speaker count. 
While the recurring presence of a single speaker across multiple IDs poses challenges for estimated variance, the critical issue for phonetic analysis is the presence of multiple speakers within a single client ID. For phonetic analysis, accurately modeling speaker variation is essential. Averaging data across distinct sources of variation can be scientifically misleading and often statistically unsound. Therefore, it is crucial to reduce speaker heterogeneity within the Common Voice client IDs if one intends to use Common Voice data for phonetic research.

Efforts have previously been undertaken to mitigate speaker heterogeneity and to acquire cleaner client IDs within the Common Voice dataset. Hintz and Siegert \cite{hintz2024commonbench} implemented a semi-automated process for cleaning client IDs, which involved the extraction of embedding features and the application of a mean- and deviation-based method to identify an initial set of outlier speakers. Subsequently, they incorporated clustering and a visualization technique, enabling human users to manually verify and refine the suspect list. Through this methodology, they eliminated 3,387 out of 15,180 client IDs.

However, this method exhibits several limitations. First, the reliance on mean and standard deviation as the primary criteria for automatic speaker recognition lacks robustness. If the utterances linked to a particular client ID consist of only a minor proportion of heterogeneous speakers, the mean and standard deviation will not be sufficiently inflated to warrant exclusion from the clean set. Conversely, the identification of a client ID with significant speaker heterogeneity does not necessitate the exclusion of all associated utterances. Employing a holistic approach that indiscriminately excludes all utterances associated with ``unclean'' client IDs may lead to substantial data loss. Furthermore, manual verification results in a bottleneck that renders scaling this approach to a larger set of client IDs impractical.

Their methodology also imposes significant constraints on the applicable data, as it is restricted to client IDs with more than 100 utterances.
Within our dataset, only a limited number of languages possess a sufficient number of client IDs exceeding 100 utterances. Specifically, only 7.97\% (8,543) of all 107,198 client IDs meet this criterion. The implementation of filtering based on \cite{hintz2024commonbench}'s approach would substantially diminish the value of Common Voice's multilingual dataset in the realm of phonetics and speech sciences.

To avoid a significant amount of data loss during the cleaning procedure, our study aims to quantify and reduce heterogeneity within the client IDs based on estimated voice characteristics and also provide an overall evaluation of the extent of heterogeneity. In addition to providing recommendations for acceptable similarity thresholds, we aim to provide users with relevant speaker comparison scores among utterances and an overall assessment such that users can make informed decisions regarding their handling of heterogeneity within the Common Voice client IDs. This approach of utterance-based cleaning would keep as much data as possible in comparison to \cite{hintz2024commonbench}'s client ID-based procedure. We use a speaker verification system (ResNet-293 trained on VoxBlink2) to obtain voice embeddings from an enrollment utterance for each client ID and all associated test utterances, quantifying voice similarity using cosine similarity. We then designed an auditing procedure to estimate a reasonable threshold for reducing heterogeneity.

We aim to answer two research questions in this study. 1) How can we quantify speaker heterogeneity within the client IDs? Vocal similarity is quantified by calculating the cosine similarity between voice embeddings from the ResNet-293 speaker verification system \cite{lin2024voxblink2}. A low similarity score should indicate that the voices of the compared utterances are less likely from the same person, thus indicating the presence of speaker heterogeneity. 
2) How can we reduce speaker heterogeneity in the client IDs? Cosine similarity alone does not automatically provide an optimal threshold for excluding utterances from heterogeneous speakers. Typically, a speaker verification system uses the ground truth and equal error rate (EER) to establish a gold threshold. Although a threshold exists for the ResNet-293 system on the English VoxCeleb1-H corpus, it might not be suitable for multilingual data. Therefore, we created a perceptual discrimination task to investigate whether the low-scoring paired utterances represent different speakers, ultimately determining a reasonable threshold across languages.

\section{Method}

\subsection{Data} 
For this study, we selected 76 languages from the Mozilla Common Voice Corpus. These languages were chosen based on their representation in the VoxCommunis Corpus \cite{ahn2022voxcommunis, zhangPacscilabVoxCommunisDatasets}. The VoxCommunis Corpus, derived from the Common Voice Corpus, includes word- and phone-level forced alignments to facilitate further phonetic analysis. It consists only of languages that have a minimum of two hours of validated utterances and a usable G2P system for transcribing the written texts into IPA symbols. Only validated utterances in Common Voice were used in the following similarity calculation. The utterances are from Common Voice versions 17 to 21, depending on when the language was included in VoxCommunis. 

\subsection{Speaker similarity estimation}

Speaker similarity is quantified using a speaker verification system. Speaker verification is a technique that determines whether to accept or reject a speaker's claimed identity by comparing a test utterance to an enrollment utterance. Two competing hypotheses are evaluated: the utterance \( U \) belongs to speaker \( S \) ($H_0$), or it does not belong to speaker \( S \) ($H_1$). These hypotheses are assessed using a speaker model, with a decision based on the likelihood ratio as expressed in Equation (\ref{eq:likelihood_ratio}).

\begin{equation}
    \Lambda = \frac{P(U|H_0)}{P(U|H_1)}
    \begin{cases} 
        \geq \tau, & \text{accept } H_0 \\ 
        < \tau, & \text{accept } H_1
    \end{cases}
\label{eq:likelihood_ratio}
\end{equation}

Here, \( P(U|H_i) \) (for \(i=0,1\)) represents the conditional probability of hypothesis \( H_i \) given utterance \( U \). The decision threshold \( \tau \) is selected based on a trade-off between the false acceptance rate (FAR) and the false rejection rate (FRR) using development data. Ideally, we would tune an automatic speaker verification system to the Common Voice data and identify issues this way. However, the Common Voice Corpus lacks Ground Truth data for accurate estimation of \( \tau \). Thus, we designed a perceptual auditing task where five authors validated speaker heterogeneity to estimate the optimal \( \tau \), as detailed in section 2.3.

This study utilized embedding vectors extracted from test utterances as speaker identity representations, denoted as \( x_e \) and \( x_t \), where cosine similarity between $x_e$ and $x_t$ was applied to obtain a speaker (or voice) similarity score also known as a verification score.
To extract the embedding feature vector, we utilized the ResNet-293 model that was pre-trained on the multilingual VoxBlink2 dataset containing 10M utterances from more than 1.1M speakers and fine-tuned on the VoxCeleb2 dataset \cite{lin2024voxblink2, nagrani2020voxceleb}. 
ResNet-293 is a deep residual neural network architecture designed for speaker verification tasks. It follows the residual learning framework where shortcut connections facilitate identity mappings, making it easier to train very deep networks and significantly increasing the depth to 293 layers to improve the model's ability to capture speaker-specific features. 

Given the diversity of languages, using a model trained on a multilingual dataset was essential as language notably influences the speaker's embedding vector \cite{2024analysis}. Furthermore, the Common Voice dataset contains significant variations in audio quality due to environmental noise, room acoustics, background speech, and music. 
The VoxBlink2 dataset, which comprises audio collected from YouTube from several languages, can address both these issues well. Pretraining on VoxBlink2 allows the model to extract speaker-specific features independent of language, and this dataset effectively captures similar acoustic variations, including reverberation and noise. Thus, training a model under the same conditions to test data ensures consistency between the training and testing data.
This strategy significantly improves generalization by first training on a massive dataset before fine-tuning on a smaller, well-structured dataset. 
Data augmentation (e.g., speed perturbation) was also used to increase variability during pre-training.

The architecture has been evaluated on large-scale speaker recognition datasets and has achieved state-of-the-art results, particularly on the VoxCeleb1-O benchmark with 0.17\% and on the VoxCeleb1-H benchmark with 0.68\% for EER \cite{lin2024voxblink2}. (The system employed by Hintz and Siegert \cite{hintz2024commonbench} for the development of CommonBench achieved a 2.9\% EER on the same VoxCeleb1-H benchmark.) These results demonstrate that ResNet-293 achieves state-of-the-art speaker verification performance, significantly outperforming previous models.

The cosine similarity criterion was then applied to the voice embeddings to compute the verification score, as defined in Equation (2):

\begin{equation}
    S(x_e, x_t) = \cos(x_e, x_t) = \frac{x_e \cdot x_t}{\lVert x_e \rVert \lVert x_t \rVert}
\label{eq:cosine}
\end{equation}
A higher score indicates a greater likelihood that the utterance belongs to the same speaker.

This system employed the final recording associated with each client ID for speaker enrollment. As a preprocessing step, client IDs with only one recording were excluded from the current study, and utterances containing fewer than three tokenized words were excluded from the evaluation. To determine the number of words in each utterance in languages that do not make use of whitespace in their orthography, we applied tokenization beforehand to obtain a reasonable estimate of this. These languages were Cantonese, Mandarin, Thai, and Japanese. For Cantonese, we used \texttt{pycantonese}; for Mandarin, \texttt{pkuseg}; for Thai, \texttt{pythainlp}; for Japanese, \texttt{fugashi} \cite{lee-etal-2022-pycantonese, pkuseg, pythainlp, mccann-2020-fugashi}. Deep models exhibit high error rates when extracting embedding vectors from such short inputs, limiting their ability to capture rich features. 
The remaining utterances from the same client ID were then compared to the enrollment utterance. A similarity score was obtained for a total of 9,204,867 test and enrollment utterance pairs.

\subsection{Validation through auditing}\label{sec:validation}
Five of the authors participated in the subsequent auditing procedure to establish a reasonable threshold of similarity score for excluding utterances. In this audit, each trial consisted of a pair of test and enrollment utterances as well. The authors labeled the trials with one of five possible choices: same speaker, different speaker, audio quality issue, missing speech, or not sure (with the instruction to use this option sparingly). The annotators were blinded to the scores of the trials while auditing. 
The full procedure was divided into two rounds. In the first round, 30 pairs were audited per language across six similarity score intervals (2,280 trials in total). In particular, five pairs were sampled from the similarity score intervals of $<0.1$, [$0.1$, $0.2$), [$0.2$, $0.3$), [$0.3$, $0.4$), [$0.4$, $0.5$), and $\geq0.5$. Each annotator then audited and labeled trials for 15 or 16 languages. In the second round, 30 trials from each annotator were sampled from the first round and audited again by the other four annotators (150 trials) to evaluate Fleiss' kappa inter-annotator consistency.

The results from the first round were then submitted to a generalized linear mixed model to identify the similarity score at which the response threshold between same and different speaker responses was equal. A subset of the data was created, containing only trials labeled as either ``same speaker'' or ``different speaker.'' The logistic regression predicted same speaker responses from a fixed effect of the similarity score, with uncorrelated random intercepts and slopes by score for annotator and language.
The crossover point when the probability is 0.5 was calculated using the formula $-\frac{\beta_0}{\beta_1}$, where ${\beta_0}$ is the model intercept and ${\beta_1}$, the coefficient for similarity score.

\section{Results}

\subsection{Similarity scores}
\begin{figure}[!htb]
    \centering
    \includegraphics[width=\linewidth]{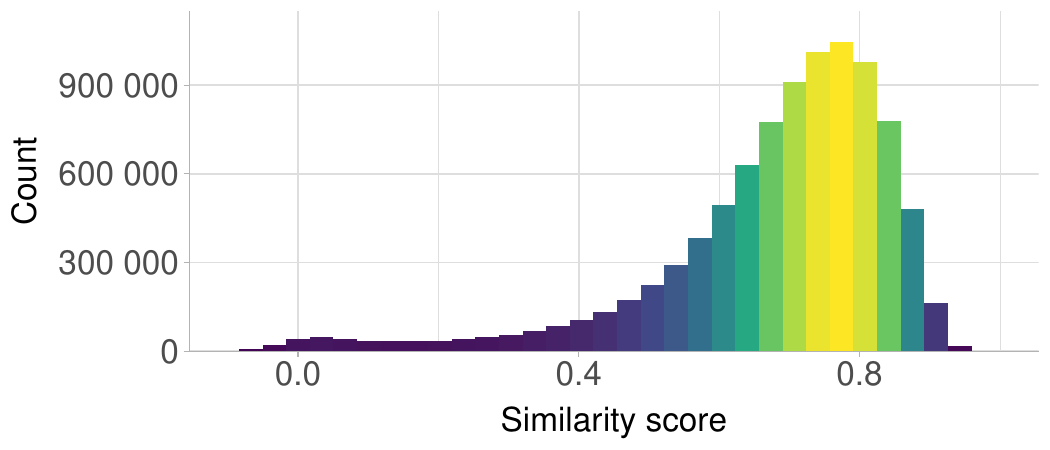}
    \caption{The distribution of the similarity scores across languages.}
    \label{fig:hist-score}
\end{figure}

The overall distribution of similarity across languages shows a strong left skew, indicating that most similarity scores are reasonably high (Figure \ref{fig:hist-score}). 
Across all languages, the inter-quartile range (IQR) of the similarity scores was 0.62 to 0.80 with a median of 0.72. 
Across individual languages, the similarity score Q1 ranged from 0.37 (Taiwanese Southern Min, nan-tw) to 0.75 (Estonian, et) with a median of 0.63, whereas Q3 ranged from 0.60 (also Taiwanese Southern Min, nan-tw) to 0.87 (also Estonian, et) with a median of 0.78. 

\subsection{Auditing results}
\begin{figure}[!htb]
  \centering
  \includegraphics[width=\linewidth]{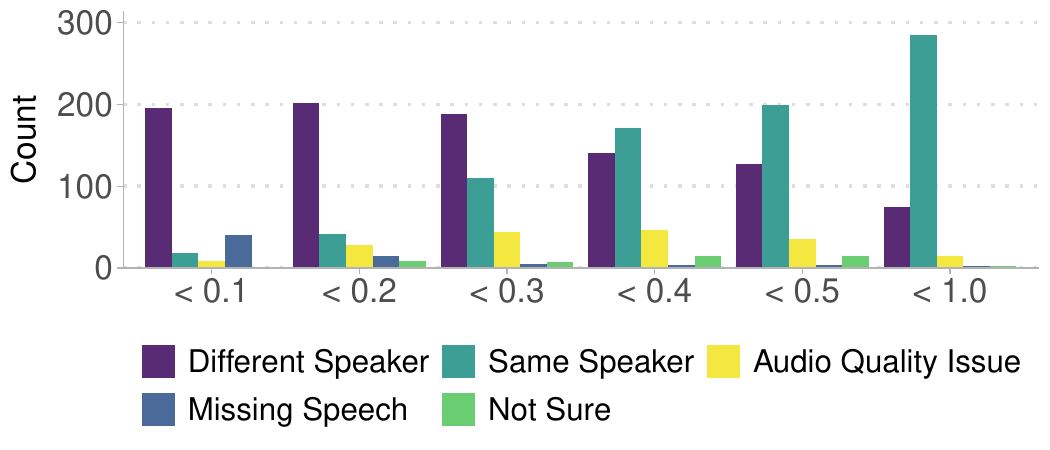}
  \caption{Counts of speaker validation audit by binned similarity score in the first round.}
  \label{fig:spkr-val}
\end{figure}

In the first round of auditing, the overall distribution of validation labels in all the trials was: same speaker (40.3\%), different speaker (45.2\%), audio quality issue (8.6\%), missing speech (3.5\%), and not sure (2.4\%). The result of the first round in each score bin is shown in Figure \ref{fig:spkr-val}. The figure shows that as the similarity score increases, the number of ``same speaker'' labels increases while the number of ``different speaker'' labels decreases, with an approximate crossover point between 0.3 and 0.4. The Fleiss' kappa inter-annotator agreement was estimated at 0.45, suggesting moderate inter-annotator consistency. 

\begin{figure}[!htb]
  \centering
  \includegraphics[width=\linewidth]{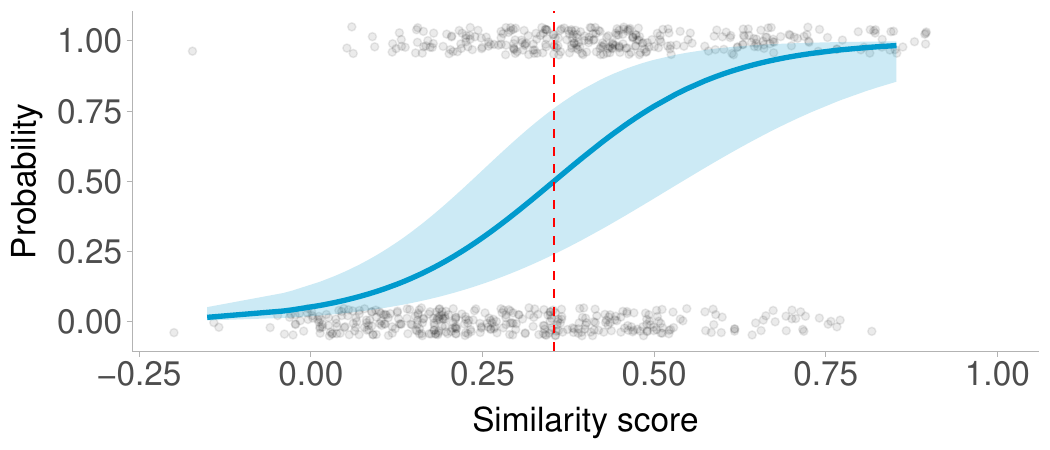}
  \caption{The distribution of same (top) and different (bottom) speaker responses, and the GLMM fit in the blue curve. The vertical line represents the similarity score (0.354) at which the probability of a same or different speaker judgment is 0.5.}
  \label{fig:glmm-result}
\end{figure}

\begin{figure}[!htb]
    \centering
    \includegraphics[width=\linewidth]{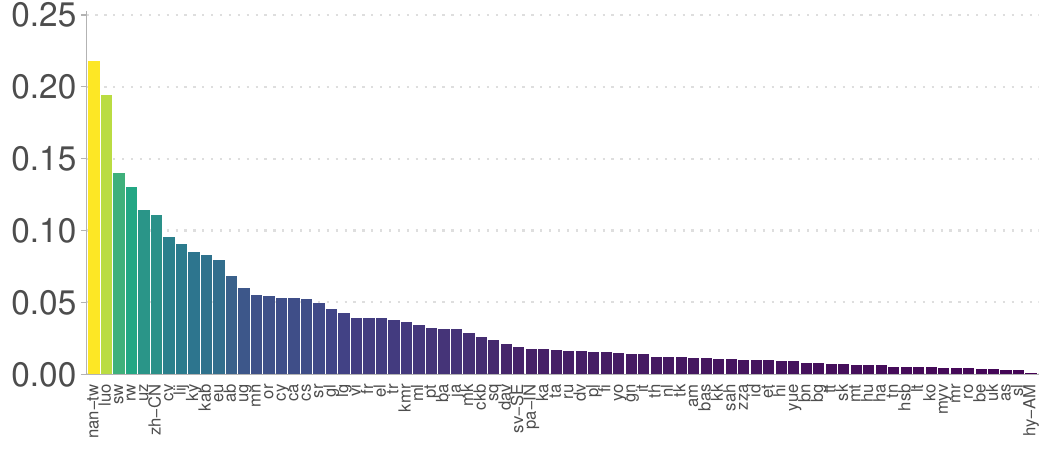}
    \caption{The proportion of utterances with a similarity score under 0.354 across languages.}
    \label{fig:dist-score}
\end{figure}

\begin{figure}[!htb]
    \centering
    \includegraphics[width=\linewidth]{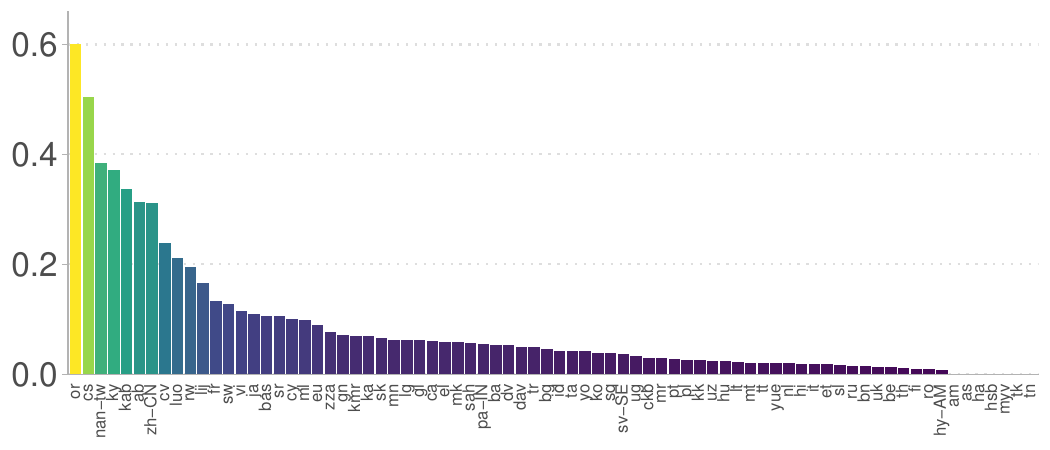}
    \caption{The proportion of client IDs that have more than 10\% of utterances with a similarity score under 0.354 across languages.}
    \label{fig:under-threshold-perc}
\end{figure}

The estimated crossover point from the logistic mixed model was at a similarity score of 0.354 (Figure \ref{fig:glmm-result}; see section \ref{sec:validation}). Using 0.354 as the threshold, we obtained the proportion data loss in each language (Figure \ref{fig:dist-score}). Across languages, the percentage data loss per language had an IQR spanning 0.8\% to 4.6\% with a median of 1.6\% and mean of 3.5\% data loss. 
Most languages (70 of 76) lost less than 10\% of the data. 
Of the 123,737 client IDs, 91.9\% (113,664) lost no more than 10\% of utterances.  
The proportion of client IDs with at least 10\% of data loss per language is shown in Figure \ref{fig:under-threshold-perc}; the IQR of this subset ranged from 2\% to 9\% of client IDs per language with a median of 4\% and a mean of 8\%.

\section{Discussion}

The goal of this paper was to quantify the speaker heterogeneity between recordings and reduce the heterogeneity within the client IDs.
To quantify heterogeneity, we used cosine similarity calculated from the ResNet-293 model pre-trained on the multilingual VoxBlink2 dataset (Figure \ref{fig:hist-score}).
To reduce heterogeneity, as we did not have the ground truth and an EER threshold of the employed ResNet-293 model was only known for English data, we carried out an auditing procedure to better understand the automatically generated speaker similarity scores.
The auditing results suggested a threshold of 0.354, which is very close to the 0.405 threshold corresponding to an EER of 0.68\% on the English-based VoxCeleb1-H corpus. Using this threshold as an approximation for mixed-speaker client IDs, we were then able to assess the degree of affected utterances and client IDs. 
The average data loss across languages using our threshold was 3.5\% of utterances, and ranged per language from $<0.0001\%$ (Armenian, hy-AM) to 21.8\% (Taiwanese Southern Min, nan-tw). 
The fact that the threshold established from the audit closely aligns with the threshold trained on English data also suggests that it is a highly reasonable and reliable metric for reducing speaker heterogeneity.

Our auditing procedure was designed to establish a crosslinguistically reasonable threshold that is effective for the multilingual dataset, rather than relying on data trained on a single language, which may introduce bias due to a lack of linguistic diversity.
In fact, since most speaker verification systems depend on models trained on monolingual data \cite{song2024introducing, chojnacka2021speakerstew, kleynhans2005language}, our approach serves as good practice when the ground truth is missing and no multilingual model is available. However, the fact that our threshold (0.354) is very close to the monolingual one (0.405) suggests that the ResNet-293 framework used in this study has the potential to be crosslinguistically robust as well, though further research is needed for verification of this.

Nevertheless, our approach has certain limitations. The five annotators achieved only a moderate level of inter-annotator consistency, indicating considerable variability in their judgments. During the second round of auditing, the annotators could not reach unanimous agreement on any of the 150 trials. This may be due to several factors. Firstly, the annotators might have diverse perspectives on what constitutes different voices from various speakers, introducing individual biases into the decision-making process. Secondly, many of the trials were sampled from languages that are non-native to the annotators. This lack of familiarity may hinder their ability to make accurate judgments, especially when prior exposure to the language is minimal (i.e., the language familiarity effect, \cite{perrachione2017speaker}).
Previous research has shown that language can also have a fairly strong influence on automatic speaker verification systems 
\cite{kleynhans2005language}. Future studies should include more annotators with diverse language backgrounds if a better threshold is desired. Additionally, by-language thresholds may be more suitable for narrowing research to a smaller group of specific languages.

However, despite the limitations mentioned, our approach is still superior to using a threshold trained on a single language or arbitrarily selecting a threshold based solely on the distribution of the data (e.g., excluding all utterances with the lowest 10\% of the score, etc.). Additionally, for other users of Common Voice datasets or the VoxCommunis Corpus, we are releasing the similarity scores from all the trials in this study to enable more informed and tailored decisions based on specific research goals and methods. Our current threshold for reducing speaker heterogeneity is merely a recommendation rather than a gold standard.

\section{Conclusion}
This study quantified and addressed speaker heterogeneity within the Mozilla Common Voice Corpus by using ResNet-based voice embeddings and an auditing procedure to establish a crosslinguistically reasonable heterogeneity threshold. Our results indicate that a threshold of 0.354 effectively reduces speaker heterogeneity while minimizing data loss. By providing speaker similarity scores and an evaluation framework, we enable researchers to make informed decisions about data usage in phonetic analysis and speech technology development. We release these scores to support further improvements in corpus phonetics and multilingual speaker verification.

\section{Acknowledgements}
This research was supported by SNF Grant PR00P1\_208460 to EC.


\bibliographystyle{IEEEtran}
\bibliography{mybib}

\end{document}